**Spin Coherence During Optical Excitation of a Single NV Center in Diamond**

G. D. Fuchs[1,2], A. L. Falk[1], V. V. Dobrovitski[3], and D. D. Awschalom[1]

1. Center for Spintronics and Quantum Computation, University of California, Santa Barbara, California, 93106

2. Cornell University, School of Applied & Engineering Physics, Ithaca, New York 14853

3. Ames Laboratory and Iowa State University, Ames, Iowa, 50011

**Abstract**

We examine the quantum spin state of a single nitrogen-vacancy (NV) center in diamond at room temperature as it makes a transition from the orbital ground-state (GS) to the orbital excited-state (ES) during non-resonant optical excitation. While the fluorescence read-out of NV-center spins relies on conservation of the longitudinal spin projection during optical excitation, the question of quantum phase preservation has not been examined. Using Ramsey measurements and quantum process tomography, we establish limits on NV center spin decoherence induced during optical excitation. Treating the optical excitation and ES spin precession as a quantum process, we measure a process fidelity of $F$=0.87±0.03, which includes ES spin dephasing during measurement. Extrapolation to the moment of optical excitation yields $F$≈0.95. This result demonstrates that ES spin interactions may be used as a resource for quantum control because the quantum spin state can survive incoherent orbital transitions.

Understanding decoherence is a fundamental pursuit in quantum information science. For solid-state "atom-like" systems that have spin, orbital, and vibronic degrees of freedom, a critical challenge is to determine which processes preserve or destroy quantum states. Nitrogen-vacancy (NV) centers in diamond have robust ground-state (GS) spin coherence at room temperature [1, 2]. Spin-orbit and spin-spin interactions are strong enough to split the sharp zero-phonon optical transition conditional on the spin state, thus enabling spin-photon entanglement and non-destructive quantum measurement of NV center spins at cryogenic temperatures [3, 4]. At room temperature, however, the direct optical transitions are broadened by phonons. Therefore, non-resonant optical excitation into the vibronic absorption band is typically used, resulting in a loss of orbital coherence between the GS and the excited state (ES) [5].

While standard fluorescence-based spin measurement of NV-centers relies on the preservation of the longitudinal (Z-axis) spin component during optical excitation [6, 7], the lack of orbital coherence at room temperature might suggest that the quantum phase of the spin could also be destroyed [8]. This issue is fundamental to understanding the non-resonant excitation dynamics of NV centers and is crucial for efforts to use coherent evolution in the ES for spin control [9, 10]. In this letter, we probe this question using Ramsey experiments where we create a coherent spin superposition in the GS, optically excite the NV center into the ES, and then probe the spin state using nanosecond-scale ES spin resonance and fluorescence measurements. The data are consistent with a theoretical model in which the transverse component of the quantum state of the NV spin is conserved during the excitation process and then decays through spontaneous emission and motional spin dephasing [10]. Treating the optical excitation combined with spin precession as a quantum process, we perform quantum process tomography (QPT) of the optical excitation and precession of a single NV spin. Our measurements indicate the process fidelity is 0.87±0.03, which is reduced by decay during measurement. When we extrapolate the measurements back to the moment of optical excitation to account for ES relaxation, the fidelity is ≈0.95.

The ground-state spin of an individual diamond NV centers can encode a qubit derived from two of the S=1 spin sub-levels at room temperature. These solid-state spins can be fabricated using ion implantation into a diamond substrate [11], and they can be coupled with nearby electronic spins [12, 13, 14] and nuclear spins [15, 16, 17] to form a local qubit register. Leveraging quantum control techniques developed for electron and nuclear magnetic resonance, NV center spins have been manipulated on sub-nanosecond time scales, enabling many operations per coherence time [18] and high-order dynamical decoupling [19]. Fast resonant manipulation also opens the door to spin control in the orbital ES of NV centers since orbital lifetimes are in the range of 10-20 ns.

To study the spin coherence during optical excitation, we prepare a spin superposition state in the orbital GS with an electron spin resonance (ESR) pulse resonant with the GS spin transition. Then, we optically excite into the ES with a precisely timed, picosecond laser pulse, followed by ES spin state measurement. To detect the spin superposition in the ES, we use nanosecond-scale ESR pulses resonant with the ES spin transition to rotate the spin state on to the Z-axis ($S_z$) for spin-selective florescence [1, 6]. The ESR pulses must be significantly faster than ES relaxation.

The carrier frequency of each ESR pulse depends on the magnetic field, *B*, applied along the NV symmetry axis. At room temperature and at large magnetic field, the spin Hamiltonians for the GS and the motionally narrowed ES are qualitatively similar, but with different values of zero-field splitting, transverse anisotropy splitting, and hyperfine splitting [9, 20, 21]. The two spin Hamiltonians are shown superimposed in Fig. 1 (a) with the $|(m_s =)0\rangle_{GS}$ and $|0\rangle_{ES}$ levels aligned at zero energy. The qubit levels we study are comprised of the $|0\rangle$ and $|-1\rangle$ electronic spin levels at *B*=1276 G. Therefore, the GS ESR pulses correspond to $f_{GS}$=0.65 GHz whereas the ES ESR pulses correspond to $f_{ES}$=2.14 GHz, indicated by arrows in Fig. 1 (a).

Figure 1 (b) outlines the optical timing sequence for the Ramsey measurement. First, we optically pump the NV center for ≈2μs with a 532 nm laser to polarize the spin into $|0\rangle_{GS}$. Optically co-aligned with this laser is the pulsed light from a tunable optical parametric oscillator (OPO) that has a pulse period of 132 ns, a wavelength of 583 nm [22] and an estimated pulse width of 5-10 ps. The experimental signal is collected using time-correlated single photon counting electronics that integrate the photons observed in a 50 ns window around each of two optical pulses after initialization. The first counting window measures the average florescence level of the NV spin in the $|0\rangle$ state after optical excitation, which is used as a reference for normalization. The second window monitors the spin-dependent fluorescence resulting from the ESR-excite-ESR pulse sequence, and is the main experimental signal. By confining light collection to these two windows associated with pulsed excitation, we select the photons relevant to the experiment. For cycles of the experiment where the NV center is not excited, we do not collect fluorescence, and hence those cycles of the experiment have no impact on the results.

The Ramsey ESR pulse sequence consists of a $\pi/2_{GS}$-pulse to rotate $|0\rangle_{GS}$ into a spin superposition state ≈20 ns before excitation into the ES with the optical pulse. After a variable delay, we apply a $\pi/2_{ES}$-pulse to map the final spin superposition onto $S_z$ for fluorescence readout. A diagram of the ESR pulses is shown in Fig. 1 (c). Both pulses are created using the direct output of an arbitrary waveform generator (AWG) operating at 19.32 GS/s with a clock referenced to the optical pulse period. This equipment enables us to generate ESR pulses of arbitrary frequencies for both the GS and ES while maintaining a fixed phase relationship between the pulses and fixed overall phase for each run of the experiment. As we delay the $\pi/2_{ES}$-pulse relative to the $\pi/2_{GS}$-pulse and the optical pulse, we keep its phase fixed, playing out exactly the same voltage signal only delayed in time. This corresponds to a Ramsey experiment performed in the "lab frame" in contrast to typical Ramsey experiments that are performed in the "rotating frame" [23]. Therefore, the Ramsey fringes we measure correspond directly to the Larmor precession rate of the ES spin.

The data shown in Fig. 2 (b) are normalized so that $P_{|0\rangle}=1$ corresponds to the florescence signal of the NV initialized to $|0\rangle_{GS}$ and $P_{|0\rangle}=0$ corresponds to the NV initialized to $|-1\rangle_{GS}$. Because $f_{ES}$ is strongly detuned from $f_{GS}$ relative to the strength of the ES driving field, the $\pi/2_{ES}$-pulse does not rotate the spin if it is applied before the optical pulse ($t_{ES}<0$). Therefore, the earliest points have $P_{|0\rangle}\approx0.5$ since the GS spin is still along the equator of the Bloch sphere (Fig. 1 (c)). As the delay increases, the Ramsey fringe amplitude increases as the $\pi/2_{ES}$-pulse crosses through the optical excitation, where it begins to rotate the spin in the ES in a coherent Ramsey measurement. Further increase of the $\pi/2_{ES}$-pulse delay causes decay of the signal envelope due to spin relaxation in the ES with characteristic rate $1/\tau^*=\Gamma+\gamma$, where Γ is spin dephasing from motional narrowing and γ is the spontaneous emission rate of the spin superposition [10]. We fit the data to a phenomenological function [24] and plot the result as a solid curve in Fig. 2 (b). From fitting we extract $\tau^*=6.0\pm0.8$ ns, the overall oscillation amplitude, $\Delta\langle S_X\rangle=0.89\pm0.04$, and the timing of optical pulse relative to the $\pi/2_{ES}$-pulse, $t_0=1.35\pm0.04$ ns. The last quantity is critical since it tells us the time of $t_{ES}=0$ in the units of our AWG clock.

The large oscillation amplitude in Fig. 2 (a) immediately suggests that the spin coherence is largely preserved through the optical excitation. The amplitude of the Ramsey fringe, $\Delta\langle S_X\rangle$, renormalized as $F = (1 + \Delta\langle S_X\rangle)/2$ = 0.95±0.04, approximates a process fidelity if we assume the longitudinal spin component is conserved and that transverse relaxation is independent of initial state. We note that this value includes dechoherence from the optical excitation and spin dephasing that occurs during the finite duration of the $\pi/2_{ES}$-pulse.

To differentiate these two contributions, we modeled the entire process using the waveforms sequenced in the experiment, along with fit values of $\tau^*$ and $t_0$ as input parameters. We also measured the spin-dependent spontaneous emission rates using time-correlated photon counting [9, 24]. The model is plotted in Fig. 2 (b) along with the data. We note that this is not a fit; there are no additional free parameters in the model, which accurately predicts the amplitude and phase. The agreement suggests that the main contribution to reduction of the fidelity is ES dephasing during the ESR pulses, and that effects from optical excitation are smaller than our experimental sensitivity.

We can characterize the evolution of arbitrary quantum states more rigorously by performing quantum process tomography (QPT) on the NV center spin with state preparation in the GS and state readout in the ES. This measurement is similar to the Ramsey measurement, except rather than varying the delay between ESR pulses, we prepare each of four initial spin states ($|+Z\rangle = |0\rangle_{GS}$, $|X\rangle = (|0\rangle_{GS} + |-1\rangle_{GS})/\sqrt{2}$, $|Y\rangle = (|0\rangle_{GS} + i|-1\rangle_{GS})/\sqrt{2}$, and $|-Z\rangle = |-1\rangle_{GS}$, and measure each along the X, Y, and Z spin axes in the ES [25]. The initial states are generated using either no pulse, a $\pi/2_{GS}$(Y)-pulse, a $\pi/2_{GS}$(-X)-pulse, or a $3\pi_{GS}$(-X)-pulse, respectively [24], after optical initialization into $|0\rangle_{GS}$, where we specify the axis of rotation in parentheses. Likewise, $S_X$, $S_Y$, and $S_Z$ spin measurements require a $\pi/2_{ES}$(Y)-pulse, a $\pi/2_{ES}$(-X)-pulse, or no pulse, immediately following optical excitation. These measurements are sensitive to the angle ($\phi$) of the spin state around the Z-axis of the Bloch sphere that evolves as $\Delta\phi = 2\pi f_{ES} t_{ES}$. Therefore, the $\pi/2_{ES}$(Y/-X)-pulse timing enables us to characterize both spin precession and decay of the quantum state due to dephasing in the ES. We also note that ESR pulse errors, which do impact our measurement, were characterized using the bootstrap tomography protocol [26]. We did not, however, apply numerical correction to the process matrix for pulse errors [24].

Using these 12 spin measurements, along with additional measurement points for signal normalization, we calculate the quantum process matrix, $\chi$. As with most experimental QPT measurements, the direct calculation of the $\chi$-matrix from experimental data ($\chi_{meas}$) yields an unphysical process due to random measurement errors. We therefore perform maximum likelihood estimation to find the closest physical $\chi$-matrix ($\chi_{phys}$) to $\chi_{meas}$ [27, 28].

The result is shown in graphical form in Fig. 3 (a), corresponding to a QPT measurement with π/2$_{ES}$-pulses with a Gaussian envelope centered at $t_0$=0.59±0.03 ns.  The large elements in the matrix correspond with real (I,I) and (Z,Z) components, along with imaginary (I,Z) and (Z,I) components.  This process matrix corresponds to a spin rotation ɸ≈90° about the Z-axis, which is precisely what we expect for spin precession, first in the GS, and then in the ES.  We note that χ$_{phys}$ characterizes all quantum processes that take place between the beginning of the GS spin preparation and the end of the ES ESR pulses. It is therefore sensitive to both the excitation process and ES spin dephasing.

The fidelity of a quantum process is defined as $F = Tr(\chi_{phys} \cdot \chi_{ideal})$, where χ$_{ideal}$ corresponds to the ideal quantum process [25].   This provides a figure-of-merit for quantum state preservation during optical excitation and allows a direct comparison with the Ramsey experiment.  From physical grounds, χ$_{ideal}$($t_{ES}$) corresponds to spin precession only.  To avoid making an assumption about the spin rotation angle, we minimized the quantity $-Tr(\chi_{phys} \cdot \chi_{ideal}(\phi))$ with respect to ɸ, where χ$_{ideal}$(ɸ) is the χ-matrix corresponding to a rotation process of angle ɸ about the Z-axis, with no decoherence.  *F* calculated with the fit value of ɸ represents the extent to which decoherence and other processes alter the quantum state from our physical expectation.  For χ$_{phys}$ in Fig. 3 (a), we find *F*=0.87±0.03. The error analysis is discussed in the supplementary information.

We also repeated the QPT measurement sequence at three additional values of $t_{ES}$ for the π/2$_{ES}$-pulses.  Since these separate measurements of the quantum process only differ by the duration of spin precession in the ES, we can use them to differentiate dephasing in the ES from loss of coherence due to the optical excitation.  The results are summarized in Fig. 3 (b), with all values of χ$_{phys}$ given in the supporting online materials.  The main plot shows the evolution of ɸ as a function of $t_{ES}$.  For clarity of the timing, we also plot the Gaussian envelope of the π/2$_{ES}$-pulses used for S$_X$ and S$_Y$ measurements.

As with the Ramsey experiments, we can also calculate what we expect to measure using the experimental ESR pulse waveforms and the independently determined input parameters.  Using these simulated measurements, we calculate a χ-matrix and see how it evolves compared to the experiment. The calculation assumes the quantum state is preserved during optical excitation and includes spin precession, finite ESR pulse duration, and dephasing in the ES from previously understood processes [10]. As with the Ramsey model calculations, the simulated QPT predicts the measured values of ϕ within experimental uncertainty.

To estimate how much of the quantum state is lost to optical excitation, we calculate *F* for the simulated QPT using the same method as the experiment. The upper left inset of Fig. 3(b) compares the simulation (purple triangles) to the experiment (blue circles). For both, *F* decreases as a function of $t_{ES}$ due to ES spin dephasing [10]. A linear guide to the eye, however, suggests that the experimental data does not extrapolate back to *F*=1 like the simulation, but instead to *F*≈0.95. This experimental estimate at $t_{ES}$=0 is consistent with our estimates of *F* from Ramsey measurements (*F*=0.95±0.04), which already account for dephasing during the ES microwave pulse. Considering that the remaining deviation from unity fidelity is on the same order as the collective uncertainties in our experiment [24], and that systematic errors generally reduce the fidelity, we regard the extrapolated *F*≈0.95 as evidence that the true process fidelity of excitation is close to *F*=1.

These measurements demonstrate that for room-temperature NV centers, spin is a robust quantum variable that largely survives incoherent optical excitation. This is a fundamental insight into NV centers' spin coherence and could advance efforts to use the ES spin Hamiltonian for rapid and coherent spin control of NV centers and nuclei. Given that incoherent excitation and spontaneous emission involve similar orbital mechanisms, these results also suggest that an NV center's full spin state is preserved during relaxation from the ES to the GS, provided the time of photon emission is accurately known.

We thank David Toyli, Christoph Weis and Thomas Schenkel for help with sample fabrication. We also thank Vladan Vuletic, Bob Buckley, and Lee Bassett for helpful discussions. We gratefully acknowledge support from AFOSR, ARO and DARPA.  Work at Ames Laboratory was supported by the Department of Energy --- Basic Energy Sciences under Contract No. DE-AC02-07CH11358.  G.D.F. also gratefully acknowledges support from Cornell University.

[1] F. Jelezko, *et al.*, *Phys. Rev. Lett*. **92**, 076401 (2004).
[2] G. Balasubramanian *et al., Nat. Maters*. **8**, 383 (2009).
[3] E. Togan *et al*., *Nature* **466**, 730 (2010).
[4] B. B. Buckley, G. D. Fuchs, L. C. Bassett, and D. D. Awschalom, *Science* **330**, 1212 (2010).
[5] K. C. Fu, *et al*. *Phys. Rev. Lett*. **103**, 256404 (2009).
[6] N. B. Manson *et al., Phys.Rev. B* **74**, 104303 (2006).
[7] L. Robeldo, H. Bernien, T. van der Sar, and R. Hanson, *N. J. Phys*. **13**, 025013 (2011).
[8] H. Uys *et al., Phys. Rev. Lett.* **105**, 200401 (2010).
[9] G. D. Fuchs *et al*., *Phys. Rev. Lett*. **101**, 117601 (2008).
[10] G. D. Fuchs *et al., Nat. Phys*. **6**, 668 (2010).
[11] D. M. Toyli *et al*., *Nano Lett*. **10**, 3168-3172 (2010).
[12] T. Gaebel *et al., Nat. Phys*. **2**, 408 (2006).
[13] R. Hanson *et al., Phys. Rev. Lett*. **97**, 087601 (2006).
[14] P. Neumann *et al, Nat. Phys*. **6**, 249 (2010).
[15] F. Jelezko *et al*., *Phys. Rev. Lett.* **93**, 130501 (2004).
[16] M. V. G. Dutt et al., *Science* **316**, 1312-1316 (2007).
[17] G. D. Fuchs, G. Burkard, P. V. Klimov, and D. D. Awschalom*, Nat. Phys*. **7**, 789 (2011).
[18] G. D. Fuchs *et al*., *Science* **326**, 1520 (2009).
[19] G. de Lange *et al*., *Science* **330**, 60, (2010); C. A. Ryan, J., S. Hodges, and D. G. Cory, *Phys. Rev. Lett*. **105**, 200402 (2010);  Naydenov *et al*., *Phys Rev. B* **83**, 081201(R) (2011).
[20] P. Neumann *et al*., *New J. Phys*. **11**, 013017 (2009).
[21] A. Batalov *et al*., *Phys. Rev. Lett*. **102**, 195506 (2009).
[22] We also performed all measurements with the OPO tuned to 553 nm, and the results were identical to those at 583 nm within our measurement uncertainty.
[23] A. Abragam, *Principles of Nuclear Magnetism* (Oxford University Press, 1961).
[24] See the supplementary online materials for details.
[25] Nielsen, M. A., and Chuang, I. L., *Quantum Computation and Quantum Information*.  (Cambridge University Press, Cambridge, 2000).
[26] V. V. Dobrovitski, G. de Lange, D. Ristè, and R. Hanson, *Phys. Rev. Lett*. **105**, 077601 (2010).
[27] J. L. O'Brien *et al*., *Phys. Rev. Lett* **93**, 080502 (2004).
[28] M. Howard *et al*., *New J. Phys*. **8**, 33 (2006).

Figure Captions

FIG. 1. (a) Spin energy-level diagram for the NV center GS and ES. Zero energy is aligned with both $|0\rangle_{GS}$ and $|0\rangle_{ES}$. There are nuclear spin sublevels in both the GS and ES (not resolved), with nuclear spin $I=1/2$ since we fabricated our NV center by ion implanting $^{15}$N ions [11]. At B=1276 G, the transition $|0\rangle_{GS} \rightarrow |-1\rangle_{GS}$ is labeled with a short blue arrow while the transition $|0\rangle_{ES} \rightarrow |-1\rangle_{ES}$ is labeled with a longer purple arrow. (b) Measurement diagram for Ramsey and QPT experiments. Optically, the NV center is initialized into $|0\rangle_{GS}$ by optical pumping with a 532 nm laser followed by waiting for fluorescence decay. The pulsed laser, which was tuned to 583 nm in the experiment, has a pulse period of 132 ns and pulse duration of 5-10 ps. For each laser pulse, there a probability ($P_{ES}$) of the NV center being excited. Using time-correlated single photon counting electronics, we bin photons collected in two windows marked with dashed black boxes. The first provides a reference for $P_{ES}$ of the initialized spin state, whereas the second forms the signal. Gaussian ESR pulses are applied before and after the optical pulse that corresponds with 'signal' collection. (c) Plot of the ESR pulses used to prepare and read out states in Ramsey and QPT experiments. The first pulse to manipulate the spin in the GS has carrier frequency $f_{GS}$=0.65 GHz, whereas the second pulse that maps the transverse spin state onto the Z-axis has carrier frequency $f_{ES}$=2.14 GHz, and is timed shortly after the laser pulse. Bloch sphere representations of the spin state indicate how the spin precession rate of the superposition state changes before and after the optical excitation.

FIG. 2. (a) Points are Ramsey data from fluorescence measurements, normalized to the fluorescence level of $|0\rangle$ ($P_{|0\rangle} = 1$) and $|-1\rangle$ ($P_{|0\rangle} = 0$). The solid red curve is a fit to the data. (b) Points are also Ramsey data from the experiment, whereas the solid red curve is the simulation, assuming no free parameters as described in the text and the supporting online materials.

FIG. 3. (a) Graphical representation of $\chi_{phys}$ for the first point of QPT with $t_{ES}$=0.59±0.03 ns. (b) Phase angle ϕ as a function of $t_{ES}$. The Gaussian envelope of the $\pi/2_{ES}$-pulses used for $S_x$ and $S_Y$ measurement in QPT are indicated above each point. Inset lower right: polar plot of ϕ, with a radius corresponding to *F* for the experiment (blue circles) and the simulation (purple triangles). The random error in ϕ due to photon shot noise is ≈5°, determined by Monte Carlo integration. The total uncertainty in ϕ is larger due to other experimental contributions (see the supplementary information). Inset upper left: *F* as a function of $t_{ES}$ for each QPT measurement (blue circles) and simulation (purple triangles). The error bars are also calculated with Monte Carlo integration using the photon shot noise statistics, and do not include other sources of uncertainty such as ESR pulse errors. The dashed lines are linear guides-to-the-eye for visual extrapolation back to the $t_{ES}$=0.

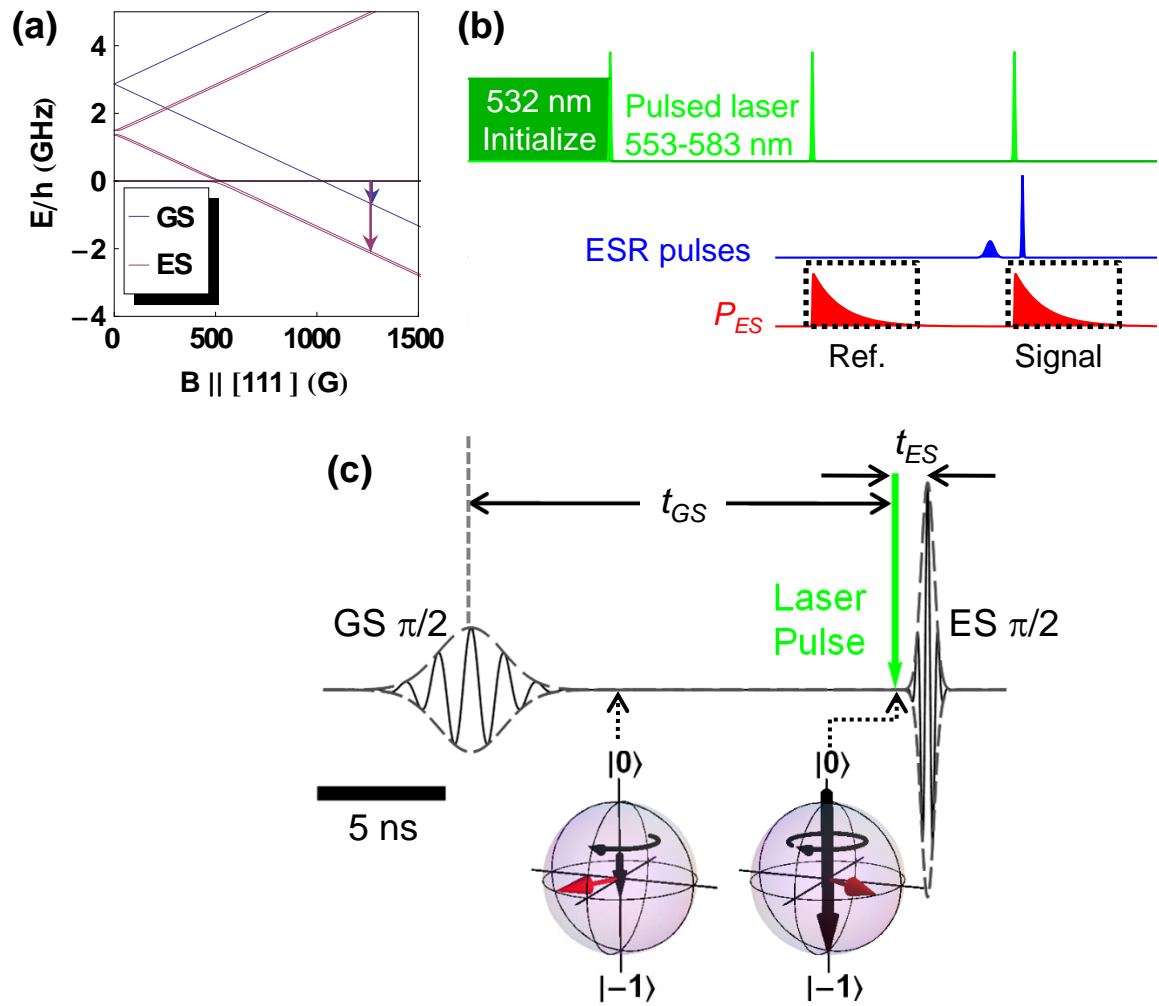

Fuchs et al., Figure 1

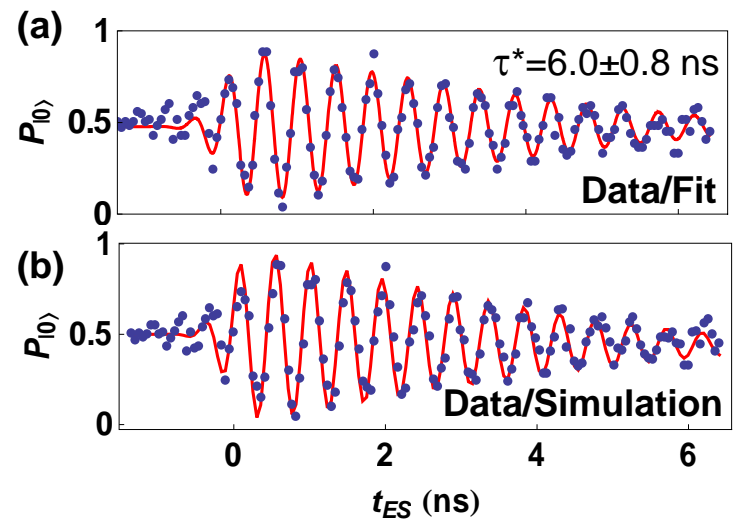

Fuchs et al., Figure 2

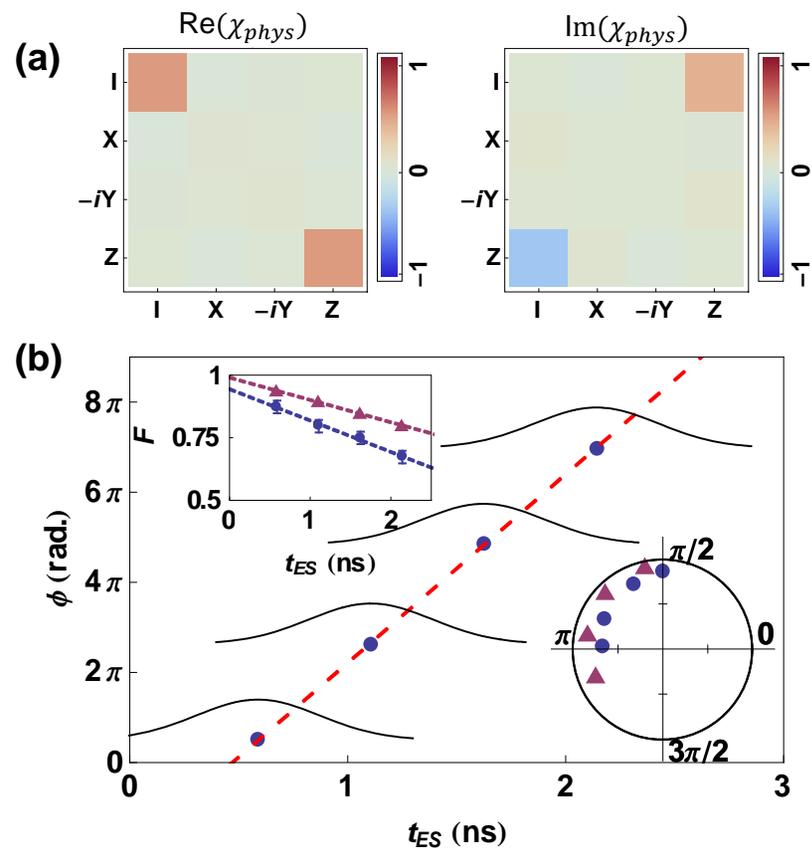

Fuchs et al., Figure 3